\documentclass[preprintnumbers,amsmath,amssymbm,prd]{revtex4}
\usepackage{epsfig}
\usepackage{graphicx}
\usepackage{amssymb}

\begin{document}

\title{Ultra-relativistic journeys through compact astrophysical objects}
%\title{A journey from pole to pole through the center of a compact star}
%\title{Traveling between the poles of a compact star: A critical-density condition for the shortest way to be the fastest way}
%\title{Two extreme trajectories between two antipodal points of a star: which one is shorter?}
%\title{Upper bound on the crossing time of a compact star}
%\title{Crossing a compact star through its center: does it take less time than orbiting it?}
%\title{Crossing a compact star through its center: in what cases will it take less time than traveling around it?}
\author{Shahar Hod}
\affiliation{The Ruppin Academic Center, Emeq Hefer 40250, Israel}
\affiliation{ } \affiliation{The Jerusalem Multidisciplinary Institute, Jerusalem 91010, Israel}
\date{\today}

\begin{abstract}

\ \ \ It has recently been proved that, for constant density stars, there is a critical 
value $\Lambda^{*}=1$ 
for the dimensionless density parameter $\Lambda\equiv 4\pi R^2\rho_{\text{max}}$ of the star 
above which the asymptotically measured travel time $T_{\text{s}}$ 
along a semi-circular trajectory that connects two antipodal points on the surface of the star 
is {\it shorter} than the travel time $T_{\text{c}}$ along the (shorter) 
straight-line trajectory that connects the two antipodal points through the center of the compact star 
[here $\{R,\rho_{\text{max}}\}$ are respectively the radius and the maximum density 
of the compact astrophysical object]. 
This intriguing observation provides a nice illustration of the general relativistic time dilation (redshift) effect 
in highly curved spacetimes. 
One expects that generic compact astrophysical objects whose dimensionless density parameters are smaller than some 
critical value $\Lambda^*$ would be characterized by the `normal' relation $T_{\text{c}}\leq T_{\text{s}}$ for 
the travel times between the two antipodal points. 
Motivated by this expectation, in the present paper we prove, using analytical techniques, 
that spherically symmetric compact astrophysical objects whose dimensionless density parameters 
are bounded from above by the model-independent 
relation $\Lambda\leq\Lambda^*={3\over2}[1-({{2}\over{\pi}})^{2/5}]$ are always (regardless of their inner 
density profiles) characterized by the normal dimensionless ratio $T_{\text{c}}/T_{\text{s}}\leq1$. 
\end{abstract}
\bigskip
\maketitle

\section{Introduction}

High school students \cite{Noteper} usually encounter the following two interesting questions during their physics studies: 
(1) What is the crossing time $T_{\text{c}}$ of a test body that falls freely 
through the center of a compact star along a straight tunnel that 
connects two antipodal points on the surface of the star \cite{Notetun}? and 
(2) What is the orbital period $T_{\text{s}}$ of a satellite that orbits a compact 
star close to its surface? 

Within the framework of Newtonian mechanics and for a {\it constant} density star of mass $M$ and radius $R$, 
the answers to these two questions are the {\it same} and can be expressed simply in terms of the 
uniform density $\rho=M/({{4\pi}\over{3}}R^3)$ of the star \cite{Noteunits}: 
\begin{equation}\label{Eq1}
T_{\text{c}}=T_{\text{s}}=\sqrt{{{3\pi}\over{4\rho}}}\  . 
\end{equation}

The two travel times $\{T_{\text{c}},T_{\text{s}}\}$ between the antipodal points of a star can be made shorter if it is assumed that the test 
particle can move arbitrarily close to the speed of light by using non-gravitational forces along its trajectories. 
In particular, in this case one finds the different traveling times (here we ignore curved spacetime effects that will 
be analyzed in detail below)
\begin{equation}\label{Eq2}
T_{\text{c}}=2R\ \ \ \ \ < \ \ \ \ \ T_{\text{s}}=\pi R\ 
\end{equation}
as measured by flat-space inertial observers. 

Within the framework of general relativity, the well known gravitational redshift effect implies that the notion of 
time is influenced by the energy density of the compact star. 
In particular, for a star with a given radius $R$, 
it is expected 
that the denser the star, the longer it takes to cross it as measured by asymptotic observers. 

Intrigued by the time dilation phenomenon in general relativity, we have recently analyzed the following 
physically interesting situation \cite{Hodrec}: Consider a physicist, Alice, who plans to send a remote-controlled spaceship between two antipodal points that are located on the surface of a compact astrophysical object. The spaceship can travel arbitrarily close to the speed of light. Alice should decide between two options:
(1) To send the spaceship along a straight-line trajectory that crosses the compact astrophysical object 
through its center \cite{Notetun}, or 
(2) To send the remote-controlled spaceship along a semi-circular trajectory on the surface 
of the compact object. 

The following important question then arises: 
Which of these two trajectories between the antipodal points of the compact astrophysical object has 
the shorter travel time as measured by the remote operator (Alice)? 

The answer to this question is quite obvious in the flat-space $M/R\to0$ limit, in which case one finds 
the `normal' inequality $T_{\text{c}}<T_{\text{s}}$ for the two travel times [see Eq. (\ref{Eq2})]. 
This inequality simply reflects the fact that, 
for highly dilute astrophysical objects, curved spacetime effects are negligible 
and the traveling times between the two antipodal points as measured 
by the remote operator are mainly determined by the (different) lengths of the two trajectories. 
 
It should be realized, however, that the general relativistic time dilation effect 
is expected to become more and more significant as the dimensionless 
compactness parameter $C\equiv {{M}/{R}}$ of the astrophysical object increases. 
In particular, using the Einstein-matter field equations, 
we have recently revealed the interesting fact that {\it constant} density stars are characterized by the 
critical value \cite{Hodrec} 
\begin{equation}\label{Eq3}
\Lambda^*=1\ \ \ \ \ \text{for constant density stars}\
\end{equation}
of the dimensionless density-area parameter $\Lambda\equiv 4\pi R^2\rho$, 
above which the travel time $T_{\text{s}}$ along 
the {\it longer} semi-circular trajectory on the surface of the compact object 
is {\it shorter} than the travel time $T_{\text{c}}$ along the 
straight-line trajectory that crosses the compact object directly through its center.

Taking cognizance of the flat-space inequality (\ref{Eq2}) and the critical relation (\ref{Eq3}) 
for constant density stars, 
it is physically reasonable to expect that, for generic astrophysical objects (that is, 
compact objects with spatially-dependent density distributions), 
there may exist a model-independent critical value $\Lambda^*$ of the dimensionless 
density parameter below which the travel time $T_{\text{c}}$ along a radial trajectory that connects 
two antipodal points of the compact astrophysical object and passes directly 
through its center is necessarily shorter than the travel time $T_{\text{s}}$ between the antipodal points along a 
semi-circular path on the surface of the compact object.

The main goal of the present paper is to derive, using analytical techniques which are based on the 
Einstein-matter field equations, a critical value $\Lambda^*$ for 
the dimensionless density parameter below which generic compact astrophysical objects are necessarily characterized 
(regardless of their inner density profiles) by the `normal' 
relation $T_{\text{c}}\leq T_{\text{s}}$ for the travel times between two antipodal points on their surfaces.  

\section{Description of the system}

We shall study, using analytical techniques, the crossing time of a spherically symmetric 
compact astrophysical object of mass $M$ and radius $R$ which is characterized by the curved line element \cite{Chan,ShTe}
\begin{equation}\label{Eq4}
ds^2=-e^{-2\delta}\mu dt^2 +\mu^{-1}dr^2+r^2(d\theta^2 +\sin^2\theta d\phi^2)\  ,
\end{equation}
where $\mu=\mu(r)$ and $\delta=\delta(r)$. 

The spatial functional behaviors of the metric functions are determined by the non-linearly coupled 
Einstein-matter field equations, $G^{\mu}_{\nu}=8\pi T^{\mu}_{\nu}$, which yield the radial 
differential equations 
\cite{May,Hodt1}
\begin{equation}\label{Eq5}
{{d\mu}\over{dr}}=-8\pi r\rho+{{1-\mu}\over{r}}\
\end{equation}
and
\begin{equation}\label{Eq6}
{{d\delta}\over{dr}}=-{{4\pi r(\rho +p)}\over{\mu}}\  .
\end{equation}
Here we have used the notations \cite{HawEl,Bond1}
\begin{equation}\label{Eq7}
\rho\equiv -T^{t}_{t}\ \ \ \ \text{and}\ \ \ \ p\equiv T^{r}_{r}
\end{equation}
for the energy density and radial pressure of the matter fields which, in the interior region of the star, 
are assumed to be non-negative. In addition, we shall assume that the matter fields inside the star 
respect the dominant energy condition \cite{HawEl,Bond1}:
\begin{equation}\label{Eq8}
0\leq p\leq\rho\ \ \ \ \ \text{for}\ \ \ \ \  r\leq R\
\end{equation}
and that they vanish,
\begin{equation}\label{Eq9}
\rho=p=0\ \ \ \ \ \text{for}\ \ \ \ \ r>R\  ,
\end{equation}
 outside the surface of the compact star.

Physically acceptable spacetimes that describe spatially regular matter configurations are 
characterized by the near-origin relations \cite{May,Hodt1}
\begin{equation}\label{Eq10}
\mu(r\to 0)\to1\ \ \ \ {\text{and}}\ \ \ \ \delta(0)<\infty\  .
\end{equation}
In addition, the metric functions of asymptotically flat spacetimes 
are characterized by the large-$r$ limits \cite{May,Hodt1}
\begin{equation}\label{Eq11}
\mu(r\to\infty) \to1\ \ \ \ {\text{and}}\ \ \ \ \delta(r\to\infty)\to0\  .
\end{equation}

Using the radial differential equation (\ref{Eq5}), one obtains the functional relation 
\begin{equation}\label{Eq12}
\mu(r)=1-{{2m(r)}\over{r}}\
\end{equation}
between the dimensionless metric function and the gravitational mass \cite{May,Hodt1}
\begin{equation}\label{Eq13}
m(r)=4\pi\int_{0}^{r} x^{2} \rho(x)dx\ 
\end{equation}
which is contained within a sphere of radius $r$. 
Taking cognizance of Eq. (\ref{Eq9}), one finds the boundary condition 
\begin{equation}\label{Eq14}
m(r=R)=M\
\end{equation}
on the surface of the compact star. 

\section{Two travel times between two antipodal points of compact astrophysical objects}

In the present section we shall analyze the travel times between two antipodal points of a compact astrophysical object for the 
two trajectories of the remote-controlled spaceship that were discussed above: 
(1) A straight-line trajectory that connects the antipodal points directly through the center of the astrophysical object, 
and (2) A semi-circular trajectory that connects the two antipodal points along the surface of the compact object. 

Our goal is to {\it minimize} the crossing time of the remote-controlled spaceship between the 
two antipodal points of the compact astrophysical object 
as measured by the asymptotically located remote operator (Alice). 
We shall therefore consider a futuristic highly advanced spaceship that may use 
non-gravitational forces which allow it to move on non-geodesic trajectories arbitrarily close 
to the speed of light. 

\subsection{The travel time along the surface of the compact astrophysical object}

For a remote-controlled spaceship that can move arbitrarily close to the speed of light, 
the travel time $T_{\text{s}}$ along a semi-circular trajectory on 
the {\it surface} of the compact astrophysical object that connects the two antipodal points 
can be obtained from the curved line element (\ref{Eq4}) of the spacetime with the properties
\begin{equation}\label{Eq15}
ds=dr=d\theta=0\ \ \ \ \  \text{and}\ \ \ \ \ \Delta\phi=2\pi\  .
\end{equation}
Substituting Eqs. (\ref{Eq12}), (\ref{Eq14}), and (\ref{Eq15}) into Eq. (\ref{Eq4}) 
and performing the azimuthal integration along 
the semi-circular trajectory, one obtains the expression
\begin{equation}\label{Eq16}
T_{\text{s}}={{\pi R}\over{\sqrt{1-{{2M}\over{R}}}}}\
\end{equation}
for the travel time between the two antipodal points along the surface of the compact object [compare the 
curved spacetime expression (\ref{Eq16}) with the naive flat-space expression (\ref{Eq2})]. 

\subsection{Generic upper bound on the crossing time through the center of a compact astrophysical object}

Assuming that the remote-controlled spaceship can move arbitrarily close to the speed of light, 
the radial crossing time $T_{\text{c}}$ of the astrophysical object as measured by the remote operator (Alice) 
can be obtained from the line element (\ref{Eq4}) of the curved spacetime with the properties
\begin{equation}\label{Eq17}
ds=d\theta=d\phi=0\  .
\end{equation}
In particular, substituting Eq. (\ref{Eq17}) into Eq. (\ref{Eq4}), one obtains the expression  
\begin{equation}\label{Eq18}
T_{\text{c}}=2\int^{R}_{0}{{e^{\delta(r)}}\over{\mu(r)}}dr\
\end{equation}
for the crossing time between the two antipodal points of the spherically symmetric astrophysical object 
along a radial trajectory that passes directly through its center \cite{Notetun}. 

We shall now derive, using {\it analytical} techniques which are based on the Einstein-matter field equations, 
a generic upper bound on the 
crossing time $T_{\text{c}}$ of the compact object as given by the integral relation (\ref{Eq18}). 
In particular, our goal is to obtain a model-independent bound on $T_{\text{c}}$ which would be valid 
for all radially-dependent density distributions that may characterize the interior regions of spatially regular 
compact astrophysical objects. 

To this end, we shall derive an upper bound on the radially-dependent metric 
function $e^{\delta(r)}/\mu(r)$ that appears in the integral relation (\ref{Eq18}). 
We first point out that Eqs. (\ref{Eq6}), (\ref{Eq8}), (\ref{Eq9}), and (\ref{Eq11}) imply 
that $\delta(r)$ is a monotonically decreasing function with the properties 
\begin{equation}\label{Eq19}
\delta(r)\geq0\ \ \ \ \ \text{for}\ \ \ \ \ r\in[0,R)\  
\end{equation}
and 
\begin{equation}\label{Eq20}
\delta(r=R)=0\  .
\end{equation}
One can therefore write the characteristic inequality
\begin{equation}\label{Eq21}
{{e^{\delta(r)}}\over{\mu(r)}}\leq\Big[{{e^{\delta(r)}}\over{\sqrt{\mu(r)}}}\Big]^2\  .
\end{equation}
 
We shall now derive a generic upper bound on the composed metric function $e^{\delta(r)}/\sqrt{\mu(r)}$ [taking cognizance of the inequality (\ref{Eq21}), one realizes that this 
bound will later help us to bound from above the function $e^{\delta(r)}/\mu(r)$ 
that appears in the expression (\ref{Eq18}) for the travel time $T_{\text{c}}$ through the center of the astrophysical object]. 
In particular, we shall bound this dimensionless function 
in terms of the dimensionless compactness parameter 
\begin{equation}\label{Eq22}
C={{M}\over{R}}\
\end{equation}
of the star and its dimensionless density-area parameter
\begin{equation}\label{Eq23}
\Lambda\equiv4\pi R^2\cdot\rho_{\text{max}}\  ,
\end{equation}
where $\rho_{\text{max}}$ is the maximum density of the astrophysical object \cite{Notermh}. 
Using the Einstein differential equations (\ref{Eq5}) and (\ref{Eq6}) with the 
functional expression (\ref{Eq12}), one finds the gradient relation
\begin{equation}\label{Eq24}
{{d\Big[{{e^{\delta(r)}}\over{\sqrt{\mu(r)}}}\Big]}\over{dr}}=-\Big[{{e^{\delta(r)}}\over{\sqrt{\mu(r)}}}\Big]\cdot
{{{{m(r)}\over{r}}+4\pi r^2p}\over{\mu r}}\
\end{equation}
for the metric function $e^{\delta(r)}/\sqrt{\mu(r)}$. 

The gravitational mass $m(r)$ contained within a sphere of radius $r\in[0,R]$ is bounded from above 
by the simple relation 
\begin{equation}\label{Eq25}
m(r)\leq {{4\pi}\over{3}}r^3\cdot\rho_{\text{max}}\  ,
\end{equation}
which yields the dimensionless inequality
\begin{equation}\label{Eq26}
\mu(r)\geq 1-{{8\pi}\over{3}}r^2\cdot\rho_{\text{max}}\ \ \ \ \ \text{for}\ \ \ \ \ r\in[0,R]\  .
\end{equation}
We shall henceforth assume that the compact astrophysical object is characterized by the 
dimensionless relation
\begin{equation}\label{Eq27}
\Lambda<{3\over2}\  ,
\end{equation}
which guarantees that the interior region of the compact object is characterized by 
the no-horizon relation [see Eqs. (\ref{Eq23}) and (\ref{Eq26})]
\begin{equation}\label{Eq28}
\mu(r)>0\ \ \ \ \ \text{for}\ \ \ \ \ r\in[0,R]\  .
\end{equation}
As we shall explicitly prove below, the assumption (\ref{Eq27}) is consistent with ({\it weaker} than) our analytically derived 
bound on the critical value of the dimensionless physical parameter $\Lambda$ which guarantees that the 
travel time $T_{\text{c}}$ through the center of the astrophysical object 
is shorter than the travel time $T_{\text{s}}$ along the surface of the compact object as measured by the 
remote operator [see, in particular, Eq. (\ref{Eq40}) below]. 

Substituting (\ref{Eq25}) and (\ref{Eq26}) into Eq. (\ref{Eq24}) and using the 
relations (\ref{Eq8}) and (\ref{Eq28}), one obtains the characteristic inequality
\begin{equation}\label{Eq29}
-{{d\Big\{\ln\Big[{{e^{\delta(r)}}\over{\sqrt{\mu(r)}}}\Big]\Big\}}\over{dr}}\leq
{{{{16\pi}\over{3}} r\rho_{\text{max}}}\over{1-{{8\pi}\over{3}} r^2\rho_{\text{max}}}}\  .
\end{equation}
Performing the integration in (\ref{Eq29}) in the radial range $[r,R]$ and using the boundary 
relations $\mu(r=R)=1-2M/R$ and $\delta(r=R)=0$ [see Eqs. (\ref{Eq12}), (\ref{Eq14}), and (\ref{Eq20})], 
one finds the relation 
\begin{equation}\label{Eq30}
\ln\Big[{{{{1}\over{\sqrt{1-2M/R}}}\cdot{{\sqrt{\mu(r)}}\over{e^{\delta(r)}}}}}\Big]\geq
\ln\Big[{{1-{{8\pi}\over{3}}R^2\rho_{\text{max}}}\over{1-{{8\pi}\over{3}}r^2\rho_{\text{max}}}}\Big]\  ,
\end{equation}
which yields the series of inequalities
\begin{equation}\label{Eq31}
{{e^{\delta(r)}}\over{\sqrt{\mu(r)}}}\leq
{{1-{{8\pi}\over{3}}r^2\rho_{\text{max}}}\over{1-{{8\pi}\over{3}}R^2\rho_{\text{max}}}}\cdot
{{1}\over{\sqrt{1-2M/R}}}\leq
{{1}\over{\big(1-{{8\pi}\over{3}}R^2\rho_{\text{max}}\big)\cdot\sqrt{1-2M/R}}}\  .
\end{equation}
From the inequalities (\ref{Eq21}) and (\ref{Eq31}) one obtains the relation
\begin{equation}\label{Eq32}
{{e^{\delta(r)}}\over{\mu(r)}}\leq
{{1}\over{\big(1-{{8\pi}\over{3}}R^2\rho_{\text{max}}\big)^2\cdot\big(1-{{2M}\over{R}}\big)}}\  .
\end{equation}

Substituting the analytically derived inequality (\ref{Eq32}) into (\ref{Eq18}) and performing the integration, 
one obtains the upper bound 
\begin{equation}\label{Eq33}
T_{\text{c}}\leq
{{2R}\over{\big(1-{{8\pi}\over{3}}R^2\rho_{\text{max}}\big)^2\cdot\big(1-{{2M}\over{R}}\big)}}\
\end{equation}
on the crossing time of the compact object along a radial trajectory that connects the two antipodal points and 
passes through its center [compare the analytically derived 
curved spacetime bound (\ref{Eq33}) with the naive flat-space expression (\ref{Eq2})]. 

\section{To cross or go around? That is the question}

In the present section we shall explicitly determine a critical value $\Lambda^*$ for the dimensionless 
density-area parameter below which the travel time $T_{\text{c}}$ of the spaceship, as measured by its 
remote operator (Alice), along a straight-line trajectory that passes 
through the center of the astrophysical object is guaranteed (for all possible matter distributions inside the object) 
to be shorter than the travel time $T_{\text{s}}$ along the semi-circular trajectory 
that connects the two antipodal points along the surface of the compact astrophysical object. 

Taking cognizance of Eqs. (\ref{Eq16}) and (\ref{Eq33}) for the travel times between the two antipodal points on the 
surface of the compact astrophysical object one concludes that, in the dimensionless regime 
\begin{equation}\label{Eq34}
C\leq{1\over2}\Big[1-{{324}\over{\pi^2(3-2\Lambda)^4}}\Big]\ \ \ \implies\ \ \ T_{\text{c}}\leq T_{\text{s}}\  ,
\end{equation}
the travel time $T_{\text{c}}$ along the straight-line trajectory is 
necessarily (that is, for all possible density profiles inside the compact astrophysical object) shorter 
than the travel time $T_{\text{s}}$ along the semi-circular trajectory: $T_{\text{c}}\leq T_{\text{s}}$.

Interestingly, using the relation (\ref{Eq34}), 
we shall now prove explicitly that a sufficient condition for the validity of the `normal' 
inequality $T_{\text{c}}\leq T_{\text{s}}$ can be formulated purely in terms of the dimensionless 
density parameter $\Lambda$ of the compact astrophysical object. 
In particular, using the characteristic inequality [see Eqs. (\ref{Eq13}), (\ref{Eq22}), and (\ref{Eq23})]
\begin{equation}\label{Eq35}
C\leq {1\over3}\Lambda\  ,
\end{equation}
one finds that if the inequality 
\begin{equation}\label{Eq36}
{1\over3}\Lambda\leq{1\over2}\Big[1-{{324}\over{\pi^2(3-2\Lambda)^4}}\Big]\
\end{equation}
is satisfied, then the sufficient condition (\ref{Eq34}) is also satisfied. 
One therefore deduces from (\ref{Eq36}) that generic astrophysical objects (that is, compact objects with generic 
density profiles) which are characterized by the dimensionless density-area relation
\begin{equation}\label{Eq37}
\Lambda\leq{3\over2}\Big[1-\Big({{2}\over{\pi}}\Big)^{2\over5}\Big]\
\end{equation}
are also characterized by the normal dimensionless ratio $T_{\text{c}}/T_{\text{s}}\leq1$ for 
the travel times between two antipodal points on their surfaces. 
It is important to emphasize the fact that the analytically derived inequality (\ref{Eq37}) for the dimensionless 
density parameter of the compact astrophysical object 
is {\it stronger} than (and therefore consistent with) the previously assumed relation (\ref{Eq27}). 

It is worth noting that the opposite 
inequality, $\Lambda>\Lambda^*={3\over2}[1-({{2}\over{\pi}})^{2/5}]$, 
for the dimensionless density-area parameter of the compact astrophysical object provides a necessary condition 
that the longer trajectory, which connects 
the two antipodal points along a semi-circular trajectory on the surface of the object, 
has a travel time $T_{\text{s}}$ which is shorter than the corresponding travel time $T_{\text{c}}$ along 
the straight-line trajectory that passes directly through the center of the astrophysical object. 

\section{Summary}

In the present compact paper we have explicitly proved that the value of the 
dimensionless density-area parameter $\Lambda\equiv 4\pi R^2\rho_{\text{max}}$ 
of a compact astrophysical object may 
determine the preferred route to be taken by a remote-controlled spaceship whose asymptotic 
operator (Alice) wants to send it in the shortest possible time \cite{Notenng}
from point A on the surface of the compact object to its antipodal point B. 

The main {\it analytical} results derived in this paper and their physical implications 
are as follows: 
%In particular, using

(1) Using the Einstein-matter field equations, 
we have derived the upper bound [see Eqs. (\ref{Eq22}), (\ref{Eq23}), and (\ref{Eq33})]
\begin{equation}\label{Eq38}
{{T_{\text{c}}}\over{2R}}\leq
{{1}\over{\big(1-{{2\Lambda}\over{3}}\big)^2\cdot\big(1-2C\big)}}\
\end{equation}
on the dimensionless crossing time of a compact astrophysical object by an ultra-relativistic spaceship 
that travels along a radial trajectory that 
passes directly through the center of the object and connects two antipodal points on its surface. 

(2) It is physically interesting to point out that one can use Eq. (\ref{Eq38}) and the characteristic inequality (\ref{Eq35}) 
in order to obtain a remarkably compact upper bound on the dimensionless 
crossing time of a compact astrophysical object which is expressed purely in terms of its 
dimensionless density parameter $\Lambda$ \cite{Notewkk}:
\begin{equation}\label{Eq39}
{{T_{\text{c}}}\over{2R}}\leq
{{1}\over{\big(1-{{2\Lambda}\over{3}}\big)^3}}\  .
\end{equation}

(3) Using the analytically derived bound (\ref{Eq38}) and the expression (\ref{Eq16}) for the travel time 
along a semi-circular trajectory on the surface of the compact object, 
we have deduced the following important conclusion [see Eq. (\ref{Eq37})]: 
\begin{equation}\label{Eq40}
\Lambda\leq\Lambda^*={3\over2}\Big[1-\Big({{2}\over{\pi}}\Big)^{2\over5}\Big]
\ \ \ \implies\ \ \ T_{\text{c}}\leq T_{\text{s}}\  .
\end{equation}
The analytically derived inequality (\ref{Eq40}) provides a sufficient condition that the 
compact astrophysical object is 
characterized by the normal dimensionless ratio $T_{\text{c}}/T_{\text{s}}\leq1$ for 
the travel times between two antipodal points on its surface as measured by the remote operator (Alice). 

Finally, we note that our Sun is characterized by the dimensionless 
relation $\Lambda\simeq7\times10^{-4}\ll\Lambda^*$ \cite{Notefu}, which immediately implies the inequality 
$T_{\text{c}}\leq T_{\text{s}}$ for the two travel times [see Eq. (\ref{Eq40})]. 
Likewise, a typical white dwarf 
with $R\simeq 4400$Km and $\rho_{\text{max}}\simeq 10^{11}$Kg/m$^3$ is characterized by the dimensionless 
relation $\Lambda_{\text{wd}}\simeq 0.018\ll\Lambda^*$ which, according to the 
analytically derived relation (\ref{Eq40}), implies the inequality $T_{\text{c}}\leq T_{\text{s}}$. 

On the other hand, a neutron star with $R\simeq 12$Km 
and $\rho_{\text{max}}\simeq 6\times10^{17}$Kg/m$^3$ is characterized by the opposite dimensionless relation $\Lambda_{\text{ns}}\simeq 0.8>\Lambda^*$, thus 
leaving open the {\it possibility} that the longer trajectory (along the surface of the star) has the shorter travel 
time as measured by the remote operator (Alice). 
In particular, in such a case one must know the exact radial profiles of the energy density and pressures inside 
the compact star in order to determine the identity of the trajectory with the shorter 
travel time between the antipodal points of the star.

%\newpage
\bigskip
\noindent {\bf ACKNOWLEDGMENTS}
%\bigskip

This research is supported by the Carmel Science Foundation. I would
like to thank Yael Oren, Arbel M. Ongo, Ayelet B. Lata, and Alona B.
Tea for stimulating discussions.


\begin{thebibliography}{99}
\bibitem{Noteper} On a personal note, I would like to add that one's first encounter with these 
interesting physical problems may occur in a homework assignment—or, as in my case, during an exam.
\bibitem{Notetun} It is assumed here that there is a thin tunnel that crosses the star through its center.
\bibitem{Noteunits} We shall use gravitational units in which $G=c=1$.
\bibitem{Notermh} Note that the constant density stars studied in \cite{Hodrec} 
are characterized by the identity $\rho_{\text{max}}=\rho$.
\bibitem{Notenng} Assuming that the remote-controlled spaceship can use non-gravitational forces, 
its velocity along a non-geodesic trajectory can in principle be arbitrarily close to the speed of light.
\bibitem{Notewkk} Note that the bound (\ref{Eq39}), 
which is expressed solely in terms of the dimensionless density parameter $\Lambda$ of the compact astrophysical object, 
is weaker than the bound (\ref{Eq38}).
\bibitem{Notefu} Note that, in terms of the physical constants $\{G,c\}$, 
the dimensionless density-area parameter of the compact astrophysical object 
is given by the expression $\Lambda\equiv 4\pi R^2\rho_{\text{max}}\cdot(G/c^2)$.
\end{thebibliography}
\end{document}